\documentclass[11pt]{article}
\usepackage{geometry}             % See geometry.pdf to learn the layout options. There are lots.
\usepackage{graphicx}
\usepackage{amssymb}
\usepackage{amsmath}
\usepackage{amsthm}
\usepackage{epstopdf}
\usepackage{comment}
\usepackage{cite}
\usepackage{color}
\usepackage[T1]{fontenc}
\usepackage{cancel,tensor,enumitem,fix-cm}
\usepackage{bbold}

\definecolor{brightpink}{rgb}{1.0, 0.0, 0.5}
\usepackage[hyperindex=true,
          pdfstartview=FitH,
          bookmarksnumbered=true,
          bookmarksopen=true,
          citecolor=blue,
          linkcolor=blue,
          colorlinks=true,
          unicode]{hyperref}

\newcommand{\be}{\begin{equation}}
\newcommand{\ee}{\end{equation}}
\newcommand{\bea}{\begin{eqnarray}}
\newcommand{\eea}{\end{eqnarray}}

\begin{document}
	
	\title{Distinguishing pure and thermal states by Landauer's principle in open systems}
	\author{
		Hao Xu
		$$\thanks{{\em E-mail }:\href{mailto: haoxu@yzu.edu.cn }
		{ haoxu@yzu.edu.cn}}
		\vspace{5pt}\\
		\small  $$Center for Gravitation and Cosmology, College of Physical Science and Technology,\\
		\small Yangzhou University, Yangzhou, 225009, China\\
	}

	\date{}
	\maketitle

\begin{abstract}{
Starting from Polchinski's thought experiment on how to distinguish between pure and thermal states, we construct a specific system to study the interaction between qubit and cavity quantum field theory (QFT) in order to provide a more operational point of view. Without imposing any restrictions on the initial states of qubit and cavity QFT, we compute the evolution of the system order by order by the perturbation method. We choose Landauer's principle, an important bound in quantum computation and quantum measurement, as the basis for the determination of the thermal state. By backtracking the initial state form, we obtain the conditions that must be satisfied by the cavity QFT: the expectation value of the annihilation operator should be zero, and the expectation value of the particle number operator should satisfy the Bose-Einstein distribution. We also discuss the difference between the thermal state and a possible alternative to the thermal state: the canonical thermal pure quantum (CTPQ) state.
}

\end{abstract}

\section{Introduction}

Prepare two sealed containers, each containing an equal amount of ice at absolute zero. We bring one of the containers into thermal equilibrium with a heat source large enough to reach a final temperature of 400K. The other container is heated with a laser to reach the same energy as the first container. The ice in both containers is now vaporized. The ice in the first container in contact with the large heat source ends up in a thermal equilibrium state, which we define as the canonical thermal state, while the ice in the second container is in a pure state. We have both containers in front of you, can you tell which is which?

\begin{figure}[h!]
	\begin{center}
		\includegraphics[width=0.8\textwidth]{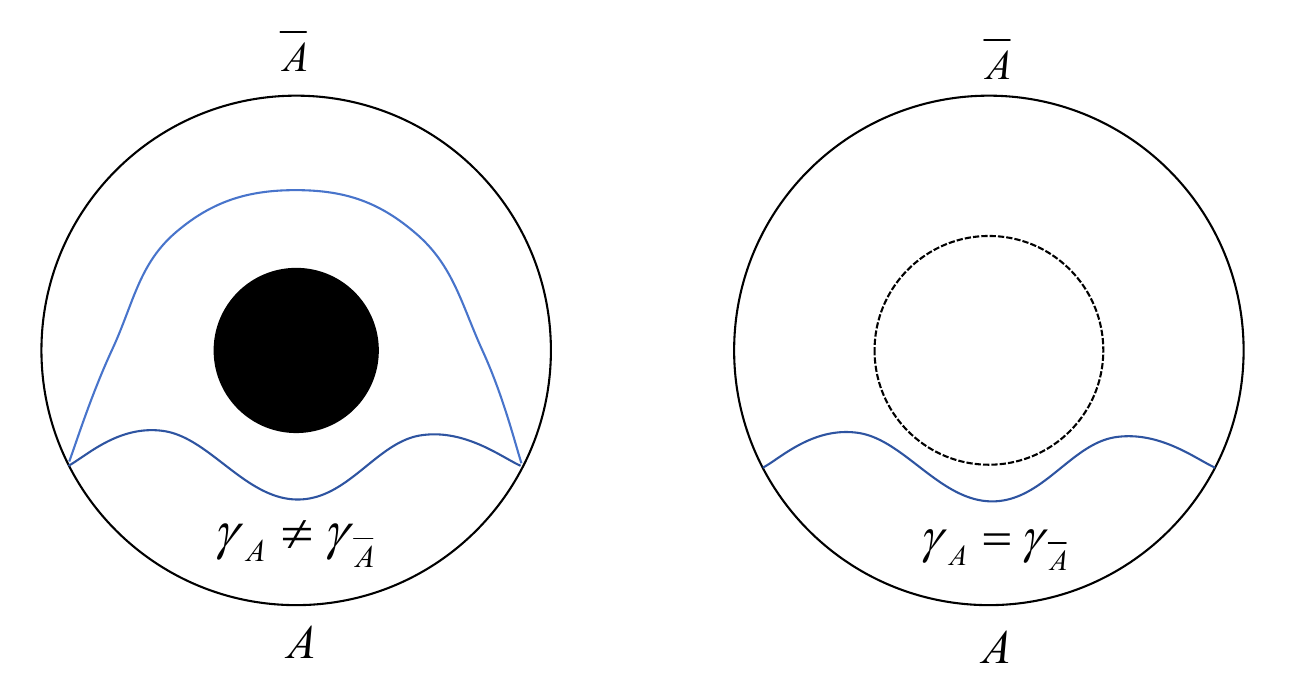}
		\vspace{-1mm}
		\caption{The minimal surfaces for the holographic calculation of entanglement entropy in the AdS spacetime. For the AdS black hole we have $\gamma_A \neq \gamma_{\bar{A}}$ (left), while in the case of black hole formation, we can find that $\gamma_A=\gamma_{\bar{A}}$ (right). This is because the collapsing shell (dashed circle) has not yet formed the event horizon. The $\gamma_A$ is the same in both cases.}
		\label{fig}
	\end{center}
\end{figure}

This is a thought experiment proposed by Polchinski in his review of the black hole information loss problem \cite{Polchinski:2016hrw}. Polchinski believes that the thought experiment may shed some light to the black hole information loss paradox, because the steam will collapse into the black hole if we heat both containers even more. In AdS/CFT we can correspond bulk physics, such as black hole formation and evaporation, to the boundary conformal field theory. For an AdS black hole, its event horizon changes the structure of spacetime. For example, when we compute the entanglement entropy of a region $A$ on the boundary field theory using the Ryu-Takayanagi formula \cite{Ryu:2006bv,Ryu:2006ef}, we find that the minimal surfaces corresponding to $A$ and the its complement $\bar{A}$ are different, and they are separated by the event horizon of the black hole. This also shows that the boundary field theory should be a mixed state. However, if the black hole has not yet formed or has evaporated, there is no event horizon. The minimal surfaces corresponds to $A$ and $\bar{A}$ coincide, and the field theory is still in a pure state. {Assuming that we are the observer on the boundary of the AdS spacetime and we use the holographic entanglement entropy as a marker for the holographic thermalization \cite{Liu:2013iza}, we would be limited to collecting information in region $A$. Although the black hole has not formed, the holographic entanglement that we use to indicate thermalization, could be the same as in the black hole case. As the observer, we will not be able to distinguish between the thermal state and the pure state, because we only have access to part of the information in the pure state.} See e.g. Figure \ref{fig}. Understanding the black hole information loss paradox requires a better knowledge of the relationship between pure and mixed states, as well as the problem of unitarity during black hole evolution \cite{Danielsson:1999fa,Takayanagi:2010wp,Roy:2015pga}.

In the review, Polchinski argues that it is impossible to distinguish between pure and mixed states. Polchinski's point is that if we know the exact form of a pure state, we can act on it directly with the projection operator to get identity, whereas acting on a mixed state gives us essentially zero. However, we do not know how the pure state is prepared, so the only thing we can do to improve accuracy is to prepare many copies and thus take multiple measurements.

From this point of view, Polchinski's conclusion is undoubtedly correct. However, when we return to an operational way of thinking about the distinction between pure and thermal states, we encounter new confusions. First, the definition of a thermal state is precise. In principle, the density matrix of the thermal state is uniquely determined whenever we are given the Hamiltonian and the temperature of the system. However, there seems to be a wider range of choices in the pure states, which is the heart of Polchinski's argument that we do not know how the pure state is prepared. This leads to confusion as to why it is so difficult for us to distinguish a rigorously well-defined system. Second, we need to define the notion of ``distinguishable''. From an operational point of view, to distinguish means to make measurements. Therefore, the theory of open quantum systems is the relevant theoretical framework for our research \cite{Zeh1970,RevModPhys.76.1267,Breuer}. The fact that two systems are indistinguishable implies that the same result is obtained for all possible physical measurements.

We now ask the question: \emph{what results do we get when we use a detector to probe separately a thermal state and a pure state that is supposedly indistinguishable from the thermal state(if such pure state exists)?} {It is equivalent to asking whether it is possible to use certain pure states as substitutes for those roles in quantum information that we usually think can only be played by thermal states. Since different motivations and system constructions may use different properties of the thermal state, this method may help us to find out exactly which operators can be used as indicators of whether a state is thermal or not in different systems, thus providing a better understanding of the significance of thermal states in physics.} 

Fortunately, we do already have an example, namely Landauer's principle \cite{landauer1961,landauer1996}. Landauer's principle relates the entropy change of a system to the heat dissipated into a reservoir during any logically irreversible computation, providing a theoretical limit of energy consumption throughout the process. It gives a direct link between information theory and thermodynamics, and establishes that \emph{information is physical}. In 2013, after a series of controversial discussions, Reeb and Wolf proposed a general and minimal setup to tighten Landauer's principle by a quantum statistical physics approach that need to be satified at the same time \cite{Reeb2013}: (i) both the ``system'' $S$ and ``reservoir'' $R$ are described by Hilbert spaces, (ii) $R$ is initially in a thermal state, (iii) $S$ and $R$ are initially uncorrelated, and (iv) the process  proceeds by unitary evolution. If all the four assumptions are satisfied, the Landauer's principle can be expressed as 
\begin{equation}
\Delta Q\geqslant {T_R}\Delta S.
\label{bound}
\end{equation}
The quantity $\Delta Q :=\text{tr}\left[{H}_R(\rho'_R-\rho_R) \right]$ is the heat transferred to the reservoir $R$, where ${H}_R$ is the Hamiltonian of $R$, while $\rho'_R$ and $\rho_R$ denote the final and initial state of $R${,} respectively. The ${T_R}$ is the temperature of $R$, and $\Delta S := S(\rho_S)-S(\rho'_S)$ is the von Neumann entropy change between the initial state $\rho_S$ and the finial state $\rho'_S$ of the system $S$.

Let us analyze the four assumptions one by one. First, both $S$ and $R$ can be described by Hilbert spaces means that both $S$ and $R$ are quantum systems. This is the easiest of the four assumptions to fulfill. Second, the initial state of $R$ is a thermal state implies that we have a strict restriction on the density matrix of the initial state of $R$, i.e., $\frac{e^{- H_{R}/{T_R}}}{\text{Tr}(e^{-H_{R}/{T_R}})}$, although there is no restriction on the form of the Hamiltonian $H_R$ and temperature ${T_R}$. Third, $S$ and $R$ are initially uncorrelated means that the density matrix of the total system in the initial state is the direct product of the density matrices of $S$ and $R$. The connection between $S$ and $R$ is established only after we introduce interactions. Fourth, the unitarity implies that the entropy of the total remains constant at all times. This is also one of the motivations for the study of Landauer's principle, that since the von Neumann entropy remains constant in unitary evolution, it can no longer be used as a measure of irreversibility. We can use the difference between the two sides of Eq.\eqref{bound}, called the entropy production, to measure irreversibility \cite{Landi:2020bsq}. If all four assumptions are satisfied, the entropy production must be non-negative. {Furthermore, we also need to emphasize that the four assumptions are just minimum settings. If extra information of the system is provided, one may be able to get a tighter bound, even in the case of $T_R=0$ that the bound \eqref{bound} becomes trivial \cite{Timpanaro2020}. However, this is not the focus of the present work, and we concentrate on the original bound \eqref{bound}.}

For the four assumptions above, we find that the second one puts a strict restriction on the form of the initial state of $R$, i.e., the thermal state. If we claim that a pure state is indistinguishable from the thermal state, it must be able to play the role of thermal state in Landauer's principle, so we can in principle just start verifying whether Landauer's principle holds when this pure state is used as a reservior. Of course, Reeb and Wolf's proof is rigorous and requires the density matrix of the thermal state {\cite{Reeb2013}}. However, to better understand the relationship between thermal and pure states, as well as other mixed states, it is helpful to consider an interaction process without specifying the initial state of the reservoirs, so that we can backtrack the form of the initial state that Landauer's principle can be achieved and then examine if there may be certain pure states that can satisfy the corresponding conditions. In the present work we will revisit the relationship between pure and mixed states in terms of system and reservoir interactions. We will choose as simple a model as possible, i.e., a qubit as the system and {a free massless bosonic scalar quantum field thoery (QFT) in a cavity} as a reservoir, with the interaction type being linear \cite{Xu:2021buk,Xu:2022juv}. Except that the initial state must be a direct product of system and reservior, we will not specifically require the initial state form for the total system, allowing us to draw more general conclusions.

The remainder of the paper is organized as follows. In section \ref{section2} we present our model and calculate the time evolution of the total system by perturbation method. We will present two theorems, which shows that in order for Landauer's principle to hold, the expectation value of the annihilation operator must be zero and the average number of particles in the cavity QFT must satisfy the Bose-Einstein distribution. In section \ref{section3} we discuss a particular state that is thought to be an alternative to the thermal state: the canonical thermal pure quantum (CTPQ) state. We examine how it relates to and differs from the thermal state. In section \ref{section4} we give a brief summary of our main results and close with conclusions.

\section{The model}\label{section2}

The total Hamiltonian $\hat{H}_{\text{total}}$ of the qubit-cavity QFT system can be written as
\begin{equation}
\hat{H}_{\text{total}}=\hat{H}_0+\hat{H}_{\text{int}},
\end{equation}
and the $\hat{H}_0$ is the sum of the free Hamiltonian of qubit and cavity QFT
\begin{equation}
\hat{H}_0=\hat{H}+\hat{H}_{\text{field}},
\end{equation}
where  $\hat{H}=\frac{\Omega}{2}{{\hat{\sigma}}_z}$ ($\hat{\sigma}_z$ is the Pauli matrix and $\Omega$ is the energy gap between ground state $|g\rangle $ and excited state $|e \rangle $ of the qubit) is the qubit Hamiltonian and $\hat{H}_{\text{field}}=\sum_{j=1}^{\infty}\omega_j \hat{a}^{\dag}_j\hat{a}_j$ is the Hamiltonian associated to the cavity QFT. The interaction Hamiltonian is given by
\begin{equation}
\hat{H}_{\text{int}}=\lambda \chi(\tau)\hat{\mu}\hat{\phi}[x(\tau)],
\end{equation}
in which $\lambda$ is a weak coupling constant so that we can apply perturbation method, $\chi(\tau)$ is the switching function that controls the interaction, $\tau$ denotes proper time, $\hat{\mu}$ is monopole operator which allows population to be exchanged between energy levels, and $\hat{\phi}[x(\tau)]$ is the field operator at the position of the qubit in the cavity.

In order to solve the time evolution of the system, we move to the interaction picture. In principle the $\hat{\mu}$ should be a linear combination of Pauli matrices. Since the $\hat{H}_0$ already contains $\hat{\sigma}_z$,  the $\hat{\sigma}_z$ in $\hat{\mu}$ is commutive with $\hat{H}_0$, so it will only produce Poincaré recurrences and does not contribute to the energy change of the qubit \cite{Hornberger2009}. Therefore, in this work we will discuss $\hat{\mu}=\hat{\sigma}_x$ as an example, and the results for $\hat{\sigma}_y$ can be obtained in a similar way. Defining $\hat{\sigma}_x :=\hat{\sigma}_+ + \hat{\sigma}_-$, which satisfies $\hat{\sigma}_+ |g\rangle=|e\rangle$ and $\hat{\sigma}_- |e\rangle=|g\rangle$, in the interaction picture we have
\begin{equation}
\hat{\mu}(\tau)=\hat{\sigma}^{+}e^{i\Omega \tau}+\hat{\sigma}^{-}e^{-i\Omega \tau},
\end{equation}
and
\begin{align}
\hat{\phi}[x(\tau)] = \sum_{j=1}^{\infty}\left( \hat{a}_je^{-i\omega_j t(\tau)} u_j\left[x(\tau)\right]+\hat{a}^{\dagger}_je^{i\omega_j t(\tau)}u_j^*\left[x(\tau)\right] \right),
\label{int}
\end{align}
where the expression of $u_j\left[x(\tau)\right]$ depends on the boundary condition.

 The time evolution operator of the system from time $\tau=0$ to $\tau=T$ is given by the Dyson series:
\begin{align}
\hat{U}(T,0)=  &\mathbb{1} \underbrace{-i\int^{T}_{0}d\tau \hat{H}_{\text{int}}(\tau)}_{\hat{U}^{(1)}} \nonumber \\ 
&\underbrace{+(-i)^2\int^{T}_{0}d\tau \int^{\tau}_{0}d\tau' \hat{H}_{\text{int}}(\tau)\hat{H}_{\text{int}}(\tau')}_{\hat{U}^{(2)}}+ ...\nonumber \\ 
&\underbrace{+(-i)^n\int^{T}_{0}d\tau ... \int^{\tau^{(n-1)}}_{0}d\tau^{(n)} \hat{H}_{\text{int}}(\tau) ... \hat{H}_{\text{int}}(\tau^{(n)})}_{\hat{U}^{(n)}},
\label{dyson}
\end{align}
so the density matrix of the total system at a time $\tau=T$ will be 
\begin{equation}
\rho_{T}\!=\!\big[\mathbb{1}+\hat{U}^{(1)}+\hat{U}^{(2)}+\mathcal{O}(\lambda^3)\big]\rho_0\big[\mathbb{1}+\hat{U}^{(1)}+\hat{U}^{(2)}+\mathcal{O}(\lambda^3)\big]^{\dagger},
\end{equation}
and we can write $\rho_T$ order by order as
\begin{equation}
\rho_{T}=\rho^{(0)}_{T}+\rho^{(1)}_{T}+\rho^{(2)}_{T}+\mathcal{O}(\lambda^3),
\end{equation}
where
\begin{align}
\rho^{(0)}_{T}&=\rho_0, \\
\rho^{(1)}_{T}&=\hat{U}^{(1)}\rho_0+\rho_0 \hat{U}^{(1)\dagger}, \\
\rho^{(2)}_{T}&=\hat{U}^{(1)}\rho_0 \hat{U}^{(1)\dagger}+\hat{U}^{(2)}\rho_0+\rho_0 \hat{U}^{(2)\dagger}.
\label{rho}
\end{align}

Tracing out the field(qubit) part we can obtain the reduced density matrix of the qubit(field). With the evolution equation in place, we also need to know the initial state of the system. For the qubit, we choose it to be 
\begin{equation}
\rho_{\text{0}}^S=\begin{pmatrix}
    p & x \\

    x & 1-p
\end{pmatrix},
\label{dq}
\end{equation}
and without loss of generality, we assume that $0<p<1/2$ and $x$ is a real number satisfying $-\sqrt{p(1-p)}\leq x \leq \sqrt{p(1-p)}$. Here the $x$ represents the coherence of the qubit. When $x = 0$, the qubit is completely decoherent, and when $x= \pm \sqrt{p(1-p)}$, the qubit is in a pure state. {Since $p+(1-p)=1$ and the range of values of $x$ ensures that the diagonal elements are both positive after doing the diagonalization, the density matrix is normalized.}

We choose the initial state for each mode of the cavity QFT to be some arbitrary state with density matrix expressed as
\begin{equation}
\rho_{\text{0}}^{f}= \bigotimes_{j=1}^{\infty} \sum_{i} c_i |\psi_i\rangle \langle \psi_i|,
\end{equation}
and for each integer $j$ we have a set of the positive coefficients $c_i$ satisfy the normalization condition $\sum_{i}c_i=1$. {The $|\psi_i\rangle$ is an arbitrary set of normalized pure states.} If for some $i$ there is $c_i = 1$ and all the rest of the coefficients are zero, then the corresponding mode is in the pure state $|\psi_i\rangle$, otherwise the mode is in a mixed state. 

The inital state of the total system is then 
\begin{equation}
\rho_{0}=\begin{pmatrix}
    p & x \\

    x & 1-p
\end{pmatrix} \otimes \bigotimes_{j=1}^{\infty} \sum_{i} c_i |\psi_i\rangle \langle \psi_i|.
\end{equation}
{The evolution is unitary, thus ensuring the normalization. Since the higher order contributions rapidly become smaller with weak coupling constant, we can analyze them order by order and find the contribution of the leading order.}

\subsection{The order of $\lambda$}

In this subsection we will prove the following theorem:

\textbf{Theorem 1}: \emph{In the order of $\lambda$, the von Neumann entropy of the qubit remains unchanged, and its energy changes just the opposite of the cavity QFT. For Landauer's principle to hold under all possible conditions, we must have $\langle \hat{a}_j \rangle:= \sum_{i} c_i \langle \psi_i |\hat{a}_j | \psi_i \rangle = 0$ for the mode $\omega_j=\Omega$, and as a result, $\Delta Q = 0$.}

\textbf{Proof}: Setting $\chi(\tau)=1$ for $0\leqslant \tau \leqslant T$, the operator $\hat{U}^{(1)}$ can be written as
\begin{equation}
\hat{U}^{(1)} = \frac{\lambda}{i}\sum_{j=1}^{\infty}(\hat{\sigma}^+\hat{a}_j^{\dagger}I_{+,j}+\hat{\sigma}^{+} \hat{a}_j I_{-,j}^{*}+\hat{\sigma}^{-}\hat{a}_j^{\dagger}I_{-,j}+ \hat{\sigma}^{-} \hat{a}_j I_{+,j}^{*}),
\end{equation}
where
\begin{equation}
I_{\pm,j}:=\int^{T}_0 d\tau~e^{i\left[\pm \Omega \tau+\omega_jt(\tau)\right]} u_j\left[x(\tau)\right].
\label{I}
\end{equation}
For any boundary condition, if the qubit is located at the same position, then $u_j\left[x(\tau)\right]$ gives a constant value. In the mode $\omega_j=\Omega$, the $I_{-,j}$ will just be proportional to the time. However, in other cases, the integration in $I_{\pm,j}$ provides only the oscillatory terms and decays rapidly as the deviation increases\footnote{If we consider quantum field theory in full space, the summation over $j$ becomes an integral of momentum, and the integral of $e^{i\left[\pm \Omega \tau+\omega_j\tau\right]}$ corresponds to the Dirac Delta function $\delta (\pm \Omega+\omega_j)$, so only the mode $\omega_j=\Omega$ contributes.}. \emph{For the remainder of this work, unless otherwise noted, we will all be discussing only the mode $\omega_j= \Omega$.}

We have
\begin{align}
\hat{U}^{(1)}\rho_{\text{in}} &= \hat{U}^{(1)}\left[ \begin{pmatrix}  
  p & x \\  
  x & 1-p  
\end{pmatrix}\otimes \sum_{i} c_i |\psi_i\rangle \langle \psi_i|\right]\nonumber \\ &= \frac{\lambda}{i} \left[\begin{pmatrix}  
  x & 1-p \\  
  0 & 0  
\end{pmatrix} \hat{a}_j \sum_{i} c_i |\psi_i\rangle \langle \psi_i| I_{-,j}^* + \begin{pmatrix}  
  0 & 0 \\  
  p & x  
\end{pmatrix} \hat{a}_j^{\dagger} \sum_{i} c_i |\psi_i\rangle \langle \psi_i| I_{-,j} \right]. 
\end{align}
Similarly
\begin{align}
\rho_{\text{in}} \hat{U}^{(1)\dagger} = -\frac{\lambda}{i} \left[\begin{pmatrix}  
  x & 0 \\  
  1-p & 0  
\end{pmatrix} \sum_{i} c_i |\psi_i\rangle \langle \psi_i| \hat{a}_j^{\dagger}I_{-,j} + \begin{pmatrix}  
  0 & p \\  
  0 & x  
\end{pmatrix} \sum_{i} c_i |\psi_i\rangle \langle \psi_i| \hat{a}_j I_{-,j}^* \right].
\end{align}
Tracing out the field part, we have the reduced density matrix of the qubit
\begin{align}
 \lambda \begin{pmatrix}  
  2x \text{Im}(\langle \hat{a}_j \rangle I_{-,j}^*)  & -i(1-2p)\langle \hat{a}_j \rangle I_{-,j}^* \\  
  i(1-2p)\langle \hat{a}_j^{\dagger} \rangle I_{-,j} & -2x \text{Im}(\langle \hat{a}_j \rangle I_{-,j}^*)  
\label{qubit1}
\end{pmatrix} ,
\end{align}
where the expectation value $\langle \hat{a}_j^{(\dagger)} \rangle$ is defined as $\langle \hat{a}_j^{(\dagger)} \rangle := \sum_{i} c_i \langle \psi_i |\hat{a}_j^{(\dagger)} | \psi_i \rangle $. Other expectation values in this work are defined in the same way as $\langle \hat{A} \rangle := \sum_{i} c_i \langle \psi_i |\hat{A}  | \psi_i \rangle $ below.

{When we calculate the von Neumann entropy corresponding to a density matrix, we have to diagonalize this density matrix first. Since we are considering a qubit, we will denote the diagonalized elements as $p_+$ and $p_-$. After diagonalization we have
\begin{equation}
\begin{pmatrix}
    p & x \\

    x & 1-p
\end{pmatrix}\rightarrow \begin{pmatrix}
    p_- & 0 \\

    0 & p_+
\end{pmatrix},
\label{dq}
\end{equation}
where
\begin{align}
p_{\pm}=\frac{1}{2}\pm \frac{1}{2}\sqrt{4p^2+4x^2-4p+1},
\label{inp}
\end{align}
and the von Neumann entropy for the initial state is
\begin{align}
S=-p_+\ln{p_+}-p_-\ln{p_-}.
\end{align}
Suppose after the interaction the density matrix becomes
\begin{equation}
\begin{pmatrix}
    p & x \\

    x & 1-p
\end{pmatrix}+
\begin{pmatrix}
    \delta p & -\delta d \\

    -\delta d^* & -\delta p
\end{pmatrix}.
\label{rhos2}
\end{equation}
Diagonalizing this density matrix we will have
\begin{equation}
\begin{pmatrix}
    p & x \\

    x & 1-p
\end{pmatrix}+
\begin{pmatrix}
    \delta p & -\delta d \\

    -\delta d^* & -\delta p
\end{pmatrix}\rightarrow \begin{pmatrix}
    p'_- & 0 \\

    0 & p'_+
\end{pmatrix}
\label{rhos2}
\end{equation}
where
\begin{align}
p'_{\pm}=\frac{1}{2}\pm \frac{1}{2}\sqrt{4p^2+4x^2-4p+1-8x\text{Re}(\delta d)+(8p-4)\delta p+4\delta p^2 +4|\delta d|^2}.
\label{finalp}
\end{align}
the von Neumann entropy for the final state is
\begin{align}
S=-p'_+\ln{p'_+}-p'_-\ln{p'_-}.
\end{align}
Comparing \eqref{inp} and \eqref{finalp} shows that the correction terms come from the $-8x\text{Re}(\delta d)+(8p-4)\delta p+4\delta p^2 +4|\delta d|^2$, where the $(-8x\text{Re}(\delta d)+(8p-4)\delta p)$ is the leading order. In the order of $\lambda$, we know the correction to the qubit is \eqref{qubit1}, so we have 
\begin{equation}
\delta p = 2 \lambda x \text{Im}(\langle \hat{a}_j \rangle I_{-,j}^*),\qquad  \text{Re}(\delta d) = -\lambda (1-2p) \text{Im}(\langle \hat{a}_j \rangle I_{-,j}^*) ,
\end{equation}
and
\begin{equation}
-8x\text{Re}(\delta d)+(8p-4)\delta p =0, 
\end{equation}
thus in the order of $\lambda$, the $p_{\pm}$ and the von Neumann entropy remain unchanged.}

Tracing out the qubit part, we have the reduced density matrix of the cavity QFT written as
\begin{equation}
\frac{\lambda}{i}x\left[ \hat{a}_j \sum_{i} c_i |\psi_i\rangle \langle \psi_i|I_{-,j}^* + \hat{a}_j^{\dagger}\sum_{i} c_i |\psi_i\rangle \langle \psi_i| I_{-,j} -\sum_{i} c_i |\psi_i\rangle \langle \psi_i|\hat{a}_j^{\dagger} I_{-j} - \sum_{i} c_i |\psi_i\rangle \langle \psi_i|\hat{a}_j I_{-,j}^*  \right],
\end{equation}
and the energy change of the field will be the trace after applying the Hamiltonian of the field to the above matrix. We have
\begin{align}
\Delta Q &=\frac{\lambda}{i}x  \omega_j [  \langle \hat{a}_j^{\dagger}\hat{a}_j \hat{a}_j   \rangle I_{-,j}^*  +\langle \hat{a}_j^{\dagger}\hat{a}_j \hat{a}_j^{\dagger} \rangle I_{-,j} -\langle \hat{a}_j^{\dagger}\hat{a}_j^{\dagger} \hat{a}_j   \rangle I_{-,j}-  \langle \hat{a}_j\hat{a}_j^{\dagger} \hat{a}_j   \rangle  I_{-,j}^*      ] \nonumber \\  
         &= -2\lambda x \omega_j \text{Im}(\langle \hat{a}_j \rangle I_{-,j}^*), 
\end{align}
where we have used the algebraic relations of the operators. Since $\omega_j= \Omega$, we can easily find the energy changes in the qubit
\begin{equation}
\text{Tr}\left[\frac{\Omega}{2}\begin{pmatrix}
    1 & 0 \\

    0 & -1
\end{pmatrix} \begin{pmatrix}
    \delta p & -\delta d \\

    -\delta d^* & -\delta p
\end{pmatrix}\right]= \Omega \delta p= 2\lambda x \Omega \text{Im}(\langle \hat{a}_j \rangle I_{-,j}^*)
\end{equation}
is just the opposite of the part of the cavity QFT.

Since the von Neumann entropy of the qubit in a $\lambda$-order perturbation remains unchanged, the right-hand side of \eqref{bound} is zero regardless of the temperature $T$ in the field theory, so we only need to have $\Delta Q \geq 0$, which implies $x \text{Im}(\langle \hat{a}_j \rangle I_{-,j}^*)\leq 0$. Since $x$ can be positive or negative, we must have $\text{Im}(\langle \hat{a}_j \rangle)= 0$.

The above analysis can also be generalized to the case $\hat{\mu}(\tau)=\hat{\sigma}_y(\tau)$. We can still define $\hat{\mu}(\tau):=\hat{\sigma}^{+}e^{i\Omega \tau}+\hat{\sigma}^{-}e^{-i\Omega \tau}$, but now the $\hat{\sigma}^{\pm}$ satisfies $\hat{\sigma}_+ |g\rangle=-i|e\rangle$ and $\hat{\sigma}_- |e\rangle=i|g\rangle$. The reduced density matrix of the qubit reads
\begin{align}
 \lambda \begin{pmatrix}  
  -2x \text{Re}(\langle \hat{a}_j \rangle I_{-,j}^*)  &(1-2p)\langle \hat{a}_j \rangle I_{-,j}^* \\  
  (1-2p)\langle \hat{a}_j^{\dagger} \rangle I_{-,j} & 2x \text{Re}(\langle \hat{a}_j \rangle I_{-,j}^*) 
\end{pmatrix}.
\end{align}
We can also have the von Neumann entropy of the qubit remains unchanged, and its energy changes just the opposite of  the cavity QFT. Now we have 
\begin{align}
\Delta Q = 2\lambda x \omega_j \text{Re}(\langle \hat{a}_j \rangle I_{-,j}^*).
\end{align}
In this case we need to have $\text{Re}(\langle \hat{a}_j \rangle) = 0$. Since $\mu(\tau)$ can be any linear combination of $\hat{\sigma}_x$ and $\hat{\sigma}_y$, in order to have $\Delta Q \geq 0$ under any condition, we can only make $\langle \hat{a}_j \rangle = 0$, which also makes $\Delta Q = 0$. { The $\delta p$ and $\delta d$ in the $\lambda$-order correction also vanish, so the term $(4\delta p^2 +4|\delta d|^2)$ is zero and would not be left to the $\lambda^2$-order. Therefore, there is no need to consider the contribution of the $\lambda$-order in the remainder of this discussion.}

$\hfill\qedsymbol$

\subsection{The order of $\lambda^2$}

In this subsection we will prove the following theorem:

\textbf{Theorem 2}: \emph{In the order of $\lambda^2$, for a positive quantity ${T_R}$, in order to have $\Delta Q \geq {T_R} \Delta S$, the initial state of the cavity QFT has to satisfy $\langle \hat{a}_j^{\dagger} \hat{a}_j \rangle := \sum_{i} c_i \langle \psi_i | \hat{a}_j^{\dagger} \hat{a}_j  | \psi_i \rangle =\frac{1}{e^{\omega_j/{T_R}}-1}$.}

\textbf{Proof}: In this case, for either $\hat{\mu}=\hat{\sigma}_x$ or $\hat{\sigma}_y$ we have
\begin{align}
& \hat{U}^{(1)} \rho_{\text {in }} \hat{U}^{(1)\dagger} \nonumber \\ 
& = \lambda^{2}\left(\hat{\sigma}^{-} \hat{a}_{j}^{\dagger} I_{-,j}+\hat{\sigma}^{+} \hat{a}_{j} I_{-,j}^{*}\right)\left(\begin{array}{cc}
p & x \\
x & 1-p
\end{array}\right) \otimes \sum_{i}c_{i}\left|\psi_{i}\right\rangle\left\langle\psi_{i}\right| \left(\hat{\sigma}^{+} \hat{a}_{j} I_{-,j}^*+\hat{\sigma}^{-} \hat{a}_{j}^{\dagger} I_{-,j}\right) \nonumber \\ 
&= \lambda^{2}\left|I_{-, j}\right|^{2}\left[\left(\begin{array}{ll}
0 & 0 \\
0 & p
\end{array}\right) \hat{a}_{j}^{\dagger} \sum_{i} c_{i}\left|\psi_{i}\right\rangle\langle\psi_{i}|\hat{a}_{j}+\left(\begin{array}{cc}
1-p & 0 \\
0 & 0
\end{array}\right) \hat{a}_{j} \sum_{i} c_{i}| \psi_{i}\rangle\left\langle\psi_{i}\right| \hat{a}_{j}^{\dagger}\right]
\end{align}
and
\begin{align}
& \hat{U}^{(2)} \rho_{\text {in }}+\rho_{\text {in }} \hat{U}^{(2) \dagger} \nonumber \\ 
&= -\frac{\lambda^{2}}{2}\left|I_{-, j}\right|^{2}\bigg[\left(\begin{array}{cc}
0 & 0 \\
x & 1-p
\end{array}\right) \hat{a}_{j}^{\dagger}\hat{a}_{j} \sum_{i} c_{i}\left|\psi_{i}\right\rangle\langle\psi_{i}|+\left(\begin{array}{cc}
p & x \\
0 & 0
\end{array}\right) \hat{a}_{j}\hat{a}_{j}^{\dagger} \sum_{i} c_{i}| \psi_{i}\rangle\left\langle\psi_{i}\right| \nonumber \\
&+\left(\begin{array}{cc}
0 & x \\
0 & 1-p
\end{array}\right) \sum_{i} c_{i}\left|\psi_{i}\right\rangle\langle\psi_{i}|\hat{a}_{j}^{\dagger} \hat{a}_{j} +\left(\begin{array}{cc}
p & 0 \\
x & 0
\end{array}\right) \sum_{i} c_{i}\left|\psi_{i}\right\rangle\langle\psi_{i}|\hat{a}_{j}\hat{a}_{j}^{\dagger}
\bigg].
\end{align}
Tracing out the field part we can have the correction to the qubit
\begin{equation}
\delta p=\lambda^2 \left|I_{-, j}\right|^{2} \left( (1-2p)\langle \hat{a}_j^{\dagger} \hat{a}_j \rangle  -p\right),
\label{p2}
\end{equation}
and
\begin{equation}
\delta d=\lambda^2 \left|I_{-, j}\right|^{2} x \left( \langle \hat{a}_j^{\dagger} \hat{a}_j \rangle +\frac{1}{2}\right).
\label{d2}
\end{equation}
Tracing out the qubit part we have
\begin{align}
&\lambda^2 \left|I_{-, j}\right|^{2}\bigg[p \hat{a}_j^{\dagger} \sum_{i} c_{i}|\psi_{i}\rangle\langle|\psi_{i}|\hat{a}_j+(1-p)\hat{a}_j \sum_{i} c_{i}|\psi_{i}\rangle\langle\psi_{i}| \hat{a}_j^{\dagger}-\frac{1-p}{2}\hat{a}_j^{\dagger} \hat{a}_j \sum_{i} c_{i}|\psi_{i}\rangle\langle\psi_{i}| \nonumber \\
&-\frac{p}{2}\hat{a}_j\hat{a}_j^{\dagger}\sum_{i} c_{i}|\psi_{i}\rangle\langle\psi_{i}| -\frac{1-p}{2}\sum_{i} c_{i}|\psi_{i}\rangle\langle\psi_{i}| \hat{a}_j^{\dagger}\hat{a}_j-\frac{p}{2}\sum_{i} c_{i}|\psi_{i}\rangle\langle\psi_{i}| \hat{a}_j\hat{a}_j^{\dagger} \bigg].
\end{align}
The energy change of the field is
\begin{align}
\Delta Q=\lambda^2 \left|I_{-, j}\right|^{2} \omega_j\left((2p-1)\langle \hat{a}_j^{\dagger} \hat{a}_j \rangle+p \right)
\end{align}
The $\Delta Q$ is independent of $x$, while $\Delta S$ is $x$-dependent. If for some constant quantity ${T_R}$ we want $\Delta Q/{T_R} \geq \Delta S$, this implies that we need $\Delta Q/{T_R}$ to be larger than the maximum value of $\Delta S$. We can find that \footnote{Here we need to insert \eqref{p2} and \eqref{d2} into \eqref{finalp} and calculate the final Von Neumann entropy, then we can have the $\Delta S$ and its derivative.}
\begin{align}
\frac{\partial \Delta S}{\partial x}=0,
\end{align}
and the value of $\Delta S$ is maximized at $x = 0$. We have
\begin{align}
 \Delta S(x=0)=-\ln{\frac{1-p}{p}}\delta p =  \lambda^2 \left|I_{-, j}\right|^{2} \ln{\frac{1-p}{p}} \left( (2p-1)\langle \hat{a}_j^{\dagger} \hat{a}_j \rangle  +p\right).
\end{align}
If
\begin{align}
(2p-1)\langle \hat{a}_j^{\dagger} \hat{a}_j \rangle+p \geq 0,
\end{align}
the $\Delta Q/{T_R} \geq \Delta S$ implies
\begin{align}
\frac{\omega_j}{{T_R}} \geq \ln{\left(\frac{1-p}{p}\right)}.
\label{ineq}
\end{align}
Combining the two equations above, if we have
\begin{align}
\frac{\omega_j}{{{T_R}}} \geq \ln{\left(\frac{\langle \hat{a}_j^{\dagger} \hat{a}_j \rangle+1}{\langle \hat{a}_j^{\dagger} \hat{a}_j \rangle}\right)},
\end{align}
it will imply
\begin{align}
\langle \hat{a}_j^{\dagger} \hat{a}_j \rangle \geq \frac{1}{e^{\omega_j/{T_R}}-1},
\end{align}
and the \eqref{ineq} must hold.

On the other hand, if $(2p-1)\langle \hat{a}_j^{\dagger} \hat{a}_j \rangle+p \leq 0$, then all the $\geq$ in the above inequalities become $\leq$ and we end up with
\begin{align}
\langle \hat{a}_j^{\dagger} \hat{a}_j \rangle \leq \frac{1}{e^{\omega_j/{T_R}}-1}. 
\end{align}
Obviously, the only one that satisfies both cases is $\langle \hat{a}_j^{\dagger} \hat{a}_j \rangle = \frac{1}{e^{\omega_j/{T_R}}-1} $, which is also known as the average number of particles in Bose-Einstein distribution.
$\hfill\qedsymbol$

\section{Discussion on Canonical Thermal Pure Quantum State}\label{section3}

Our question now is which type of initial state satisfies the two theorems proved in section \ref{section2}. The thermal state definitely matches the requirements. We can write the density matrix of the thermal state as \cite{Olivares2012}
\begin{equation}
\bigotimes_{j=1}^{\infty}\sum_{n_j=0}^{\infty}\frac{\bar{n}_j^{n_j}}{(1+\bar{n}_j)^{1+n_j}}|n_j\rangle\langle n_j|,
\label{thermal}
\end{equation}
where for each integral value $j$, $n_j \in [0,\infty)$, and $\bar{n}_j:={1}/{\left(e^{\frac{\omega_j}{{T_R}}}-1\right)}$. 
Since \eqref{thermal} contains only diagonal terms, we have $\langle \hat{a}_j \rangle = 0$, and for each $j$
\begin{equation}
\langle \hat{a}_j^{\dagger} \hat{a}_j \rangle = \sum_{n_j=0}^{\infty}\frac{\bar{n}_j^{n_j} n_j}{(1+\bar{n}_j)^{1+n_j}} = \bar{n}_j.
\end{equation}
The result corresponds exactly to Theorem 1 and Theorem 2  we proved.

So is a pure state possible? One idea is to construct the following kind of pure state, called canonical thermal pure quantum (CTPQ) state(unnormalized) \cite{Sugiura2012,Sugiura2013,Sugiura2014}
\begin{equation}
|\beta,\hat{H} \rangle=\sum _i z_i \exp{\Big(-\frac{\beta \hat{H}}2}\Big)|i\rangle,
\end{equation}
where $z_i$ is a set of random complex numbers, whose real and imaginary parts satisfy the unit normal distribution, $\beta$ is the inverse temperature, and $|i\rangle$ is a set of orthogonal normalized bases in a Hilbert space. We have
\begin{align}
\langle \hat{a}_j \rangle&=\frac{\sum_{k=0}\langle k|\exp{}(-\beta \omega_j k/2)z_k^* \hat{a}_j\sum_{i=0}z_i\exp{(-\beta \omega_j i/2)}|i\rangle }{\sum_{i=0}|z_i|^2\exp{(-\beta \omega_j i)}} \nonumber \\
&=\frac{\sum_{k=0}\langle k|\exp{}(-\beta \omega_j k/2)z_k^* \sum_{i=0}z_i\exp{(-\beta \omega_j i/2)}\sqrt{i}|i-1\rangle }{\sum_{i=0}|z_i|^2\exp{(-\beta \omega_j i)}}\nonumber \\
&=\frac{\sum_{i=0}z_i^* z_{i+1}\sqrt{i+1}\exp{(-\beta\omega_j(2i+1)/2)} }{\sum_{i=0}|z_i|^2\exp{(-\beta \omega_j i)}}
\end{align}
and
\begin{align}
\langle \hat{a}_j^{\dagger}\hat{a}_j \rangle=\frac{\sum_{i=0}|z_i|^2\exp{(-\beta \omega_j i)}i }{\sum_{i=0}|z_i|^2\exp{(-\beta \omega_j i)}}.
\end{align}
For an arbitrary set of $z_i$-sequences, the above result does not support the Theorems 1 and 2. This is because even if we are taking values with unit normal distribution, one may have the stochastic error from $z_i$ that cannot be ignored. However, if we take enough sets of $z_i$ and then average them, we will have $\overline{z_i^* z_{i+1}}=0$ and $\overline{|z_i|^2} = 1$, thus satisfying the conditions.

It is necessary to emphasize the difference between the CTPQ state and the thermal state. It is well known that the pure state is a vector in a Hilbert space, while the mixed state is not. If we compute the expectation value of an operator $\hat{A}$ in some mixed state $c_i |\psi_i\rangle \langle \psi_i|$, we have
\begin{equation}
\langle \hat{A} \rangle = \text{Tr}\left( c_i \langle \psi_i| \hat{A} |\psi_i\rangle\right).
\end{equation}
Here we have to do two averages. The first is to find the expectation value of the operator $\hat{A}$ on the pure state $|\psi_i\rangle$, since the state $|\psi_i\rangle$ may not be the eigenstate of the operator $\hat{A}$. The second is to evaluate the distribution contributed by the coefficient $c_i$.

For the pure state, there is no second averaging, since it is a definite vector in Hilbert space. The CTPQ state, on the other hand, does not choose a definite vector at the beginning, but mixes in some randomness. By averaging multiple random vectors again, it has the same results as the thermal state. For most systems in nature there are typically a large number of particles and the dimension of a Hilbert space grows exponentially, and the averages of mean and variance of the operator will converge to the thermal result \cite{Sugiura2017}. But we should also note that this similarity is based on probability, and when the particle number is not large enough or we take the value in a single run, the final result may still be very different from the thermal state.

In our model, although there are infinite modes in the cavity QFT, only one mode actually contributes to the result. Theorems 1 and 2 strictly hold when this mode is in a thermal state, but if we construct a pure state, theorems 1 and 2 may not hold because the randomization itself introduces errors. We can get the result of a thermal state by selecting multiple pure states and then averaging them, but this would probably no longer be the true definition of a pure state.

\section{Conclusions and outlook}\label{section4}

In this work, we revisit the thought experiment proposed by Polchinski on whether pure and thermal states can be distinguished. Since the thermal state appears as a necessary condition in Landauer's principle, we construct a model to study the interaction between the qubit and the cavity QFT, where the qubit serves as the system $S$ and the cavity QFT serves as the reservoir $R$. To better understand the significance of the thermal state, we did not specify the initial state of the cavity QFT, but rather asked what conditions the cavity QFT would have to satisfy if Landauer's principle holds. In a sense, this work is not written to discover new things, but to explain what we already know in a new way. That is, assuming that certain results hold, to use a operational way to trace back to what we need to find, so that we can understand which operators can be used as indicators of the thermal state. This may lead to a better understanding of certain physical conditions. Using the perturbation method, we obtain two theorems:
\begin{itemize}
  \item \emph{In the order of $\lambda$, the von Neumann entropy of the qubit remains unchanged, and its energy changes just the opposite of the cavity QFT. For Landauer's principle to hold under all possible conditions, we must have $\langle \hat{a}_j \rangle:= \sum_{i} c_i \langle \psi_i |\hat{a}_j | \psi_i \rangle = 0$ for the mode $\omega_j=\Omega$, and as a result, $\Delta Q = 0$.}

\item \emph{In the order of $\lambda^2$, for a positive quantity ${T_R}$, in order to have $\Delta Q \geq {T_R} \Delta S$, the initial state of the cavity QFT has to satisfy $\langle \hat{a}_j^{\dagger} \hat{a}_j \rangle := \sum_{i} c_i \langle \psi_i | \hat{a}_j^{\dagger} \hat{a}_j  | \psi_i \rangle =\frac{1}{e^{\omega_j/{T_R}}-1}$.}
\end{itemize}

These two theorems are valid for the thermal state. Do they hold for the pure state? We discuss one possible candidate, the CTPQ state. It is defined as a set of orthogonal normalized bases in a Hilbert space with the operator $\exp{\big(-\frac{\beta \hat{H}}2}\big)$ acting on them, and mixed in some random numbers. Averaging these random numbers gives the same result as the thermal state. When the number of particles in the system is large, the dimensions of the Hilbert space also become large, and the average errors generated by random numbers become smaller. It {is} probably more like a mixed state, but it still gives us a good clue to explore the difference between the pure and mixed states.

Returning to the original question, can a pure state be operationally indistinguishable from a thermal state? Our answer is it depends on the specific system and the definition of the pure state. For most of the systems in nature, there are typically a large number of particles and the dimension of a Hilbert space. The thermal state corresponds to a specific distribution. For the pure state, if all particles are in a particular pure state, or the total system is in a pure state, it may not give the same result as the thermal state. Its similarity to the thermal state is based on probability. However, if each particle is in a random pure state seperately, or we are considering a large number of pure systems as the ensemble, using a detector to interact with them may obtain the same result as the thermal state.

We conclude with a brief discussion of possible applications of this work. First, we can return to the more fundamental physics problem of how the thermal state and temperature are actually defined. If we consider a large system that is in a microcanonical distribution and the internal interactions are weakly coupled, the temperature emerges as a parameter that fixes the equilibrium condition, and the canonical partition function can be obtained by saddle-point approximation \cite{Martin1959}. The canonical thermal system can be considered as a subsystem of the microcanonical system, and this allows us to work on the definition and properties of the thermal state from the source, e.g. by considering stronger couplings \cite{Moreno2019,Xu:2021ihm} or the genalization of microcanonical systems \cite{Iyoda2017}. {In addition, it has also been recently shown that even when idealized measurements cannot be made, it is still possible to read out the symmetric characteristic function of the states of QFT by using state tomography \cite{Tian:2023sfz}. All of these directions may be the subject of future research.}

Second, the qubit in this work is at a fixed position in an inertial reference system. We can also consider the cases of accelerating detectors \cite{Unruh:1976db,Unruh:1983ms} and detectors in curved spacetime \cite{Davies,Hodgkinson:2014iua,Ng:2016hzn}, which will help us further understand the quantum information in non-inertial systems. For example, in this work the variation of the off-diagonal elements of the qubit corresponds to the decoherence effect, and there are also some recent literatures discussing the connection between decoherence and accelerating detectors \cite{Nesterov:2020exl,Xu:2023tdt}. The accelerating detectors may also be able to shed some light on the problem of quantum gravity \cite{Alkofer:2016utc}. {Furthermore, in this work we consider bosonic field, so the Bose-Einstein distribution is satisfied. For fermionic field, we believe the results must be different. A discussion of Landauer's principle for fermionic field can be found in \cite{Cao:2023hwu}. We can also consider different field theories, especially the conformal field theory and effective field theory of gravity \cite{Burgess:2003jk}.}

Finally, Polchinski's motivation for proposing the thought experiment was to better understand the black hole information loss paradox, so investigating the time evolution of the CFT at the boundary corresponding to the matter collapse process in the bulk is a more straightforward choice \cite{Anous:2016kss}. It would be interesting to try to compute observables of this process, such as the two-point correlation function, and compare them with the results in AdS/CFT. We will pursue these directions in our future work.

\section*{Acknowledgements}
Hao Xu thanks National Natural Science Foundation of China (No.12205250) for funding support.

%\bibliographystyle{utcaps}
%\bibliography{../../../BibLibrary/papers2.bib,../../../BibLibrary/papers1.bib}
%\end{document}

\providecommand{\href}[2]{#2}\begingroup%\raggedright
\small\itemsep=0pt
\providecommand{\eprint}[2][]{\href{http://arxiv.org/abs/#2}{arXiv:#2}}

\end{document}